# Quantum and classical statistics of the electromagnetic zero-point field

Michael Ibison

Institute for advanced studies at Austin

4030 Braker Lane West, suite 300, Austin, TX 78759

ibison@earthtech.org

Bernhard Haisch

Solar and Astrophysics Laboratory

Div. 91-30, Bldg. 252, Lockheed Martin, 3251 Hanover St., Palo Alto, CA 94304

haisch@sag.space.lockheed.com

# ABSTRACT

A classical electromagnetic zero-point field (ZPF) analogue of the vacuum of quantum field theory has formed the basis for theoretical investigations in the discipline known as random or stochastic electrodynamics (SED). In SED the statistical character of quantum measurements is imitated by the introduction of a stochastic classical background electromagnetic field. Random electromagnetic fluctuations are assumed to provide perturbations which can mimic certain quantum phenomena while retaining a purely classical basis, e.g. the Casimir force, the Van-der-Waals force, the Lamb shift, spontaneous emission, the RMS radius of a quantum-mechanical harmonic oscillator, and the radius of the Bohr atom. This classical ZPF is represented as a homogeneous, isotropic ensemble of plane electromagnetic waves whose amplitude is exactly equivalent to an excitation energy of hv/2 of the corresponding quantized harmonic oscillator, this being the state of zero excitation of such an oscillator. There is thus no randomness in the classical electric field amplitudes: Randomness is introduced entirely in the phases of the waves, which are normally distributed. Averaging over the random phases is assumed to be equivalent to taking the ground-state expectation values of the corresponding quantum operator. We demonstrate that this is not precisely correct by examining the statistics of the classical ZPF in contrast to that of the electromagnetic quantum vacuum. Starting with a general technique for the calculation of classical probability distributions for quantum state operators, we derive the distribution for the individual modes of the electric field amplitude in the ground-state as predicted by quantum field theory (QFT). We carry out the same calculation for the classical ZPF analogue, and show that the distributions are only in approximate agreement, diverging as the density of **k** states decreases. We then introduce an alternative classical ZPF with a different stochastic character, and demonstrate that it can exactly reproduce the statistics of the electromagnetic vacuum of QED. Incorporating this field into SED, it is shown that the full probability distribution for the amplitude of the ground-state of a quantum-mechanical harmonic oscillator can be derived within a classical framework. This should lead to the possibility of developing further successful correspondences between SED and QED.

# I. INTRODUCTION

The research program known as random or stochastic electrodynamics (SED) is clearly described in the classic review article by Boyer [1] and in the recent monograph of Milonni [2]. SED is basically a modern extension of the classical electron theory of Lorentz and a follow-on to investigations of Planck [3], Nernst [4] and Einstein and Stern [5], but with a different choice of boundary conditions consistent with relativity; the boundary conditions being discussed in detail by Boyer [1]. The uniform, homogeneous zero-point field (ZPF) forming the basis of SED today has a Lorentz-invariant radiation spectrum, i.e. the energy density scales as $\rho(\nu) \propto \nu^3$. Planck's constant, $h$, appears in the electromagnetic field strength but only in the role of a scale factor. In the classical SED view, quantum fluctuations are equivalent to electromagnetic ZPF perturbations, and thus the strength of the electromagnetic field must involve $h$; however whereas in the quantum view $h$ is a unit of quantization, in the SED view $h$ is merely a measure of field strength. The primary emphasis of this research program has been to determine to what extent classical physics plus a ZPF can reproduce the results of quantum mechanics (QM) and quantum electrodynamics (QED).

Several interesting classical re-workings of quantum problems have been published. For example, it has been shown by Boyer [6] that the blackbody spectrum can be derived in this fashion without recourse to quantum assumptions. Inherent discontinuity in available energy states is not required in this approach since the same net effect is produced on an ideal oscillator by the random fluctuations of the ZPF. Another classical problem, the stability of the Bohr atom, similarly has a different resolution from the perspective of SED. A classically orbiting electron in the sense of the Rutherford planetary model of 1913 would pick up energy from the ZPF at a rate exactly balancing the radiative losses due to accelerated motions of the charged electron [1][7]. This balance has been demonstrated only for the ground-state of the Bohr hydrogen atom.

As intriguing as these and other quantum re-derivations are, a fundamental objection to the overall approach of SED is that if the ZPF is regarded as a real electromagnetic field, that would have unacceptable gravitational consequences since the energy density would presumably give rise to a physically unrealistic cosmological constant. Recent SED studies address this problem and, building on a conjecture of Sakharov [8], find a possible interpretation of gravitation and inertia as themselves ZPF-mediated phenomena [9][10], analogous to the view of ZPF-mediated quantum phenomena. Some support for the possibility of non-gravitating vacuum energy has recently been proposed by Dutta [11]. Nevertheless, the SED concepts have met with justifiable skepticism given the limitations of SED vis-à-vis modern quantum theory. We address one such limitation in this paper.

While SED is suggestive of interesting physics, given the resounding success of quantum theory as a predictive description of nature, it will be necessary to demonstrate a far more detailed correspondence of SED with QFT. Indeed, we have identified a discrepancy between SED as presently formulated and QFT in that the ZPF is assumed to consist of electromagnetic fields that are perfectly random in phase but with a known amplitude, which differs from the distribution for the individual modes of the electric field amplitude in the ground-state as predicted by QFT. We introduce an alternative classical ZPF with a different stochastic character, and demonstrate that it exactly reproduces the statistics of the electromagnetic vacuum of QED. Incorporating this field into SED, it will be shown that the full probability distribution for the amplitude of the ground-state of a quantum-mechanical harmonic oscillator can be derived within a classical framework.

Our approach is as follows. In section II we present an elementary illustration of the problem under consideration. In section III we first introduce some notation leading to a definition of an equivalent classical probability distribution for a quantum state operator. Then we derive the ground-state probability distributions for the full electric field and a single mode of the electric field consistent with QFT. In section IV SED distributions for each of these are computed and compared with those of QFT. In section V the computation and comparison is repeated using our modified SED. We then apply the modified SED field to determination of the probability distribution for the position of a classical harmonic oscillator and compare with the standard quantum mechanical result, demonstrating precise agreement. This modification of the standard SED approach should clear the way for further development of correspondences with QFT.

## II. CLASSICAL VS. QUANTUM SIMPLE HARMONIC OSCILLATOR

The discrepancy between SED and QFT that we seek to address can be illustrated by examining a simple harmonic oscillator (cf. Davies and Betts [12]). For a mass *m*, on a spring with force constant K, the frequency of oscillation will be $\omega = \sqrt{K/m}$ and the position x(t) can be expressed as

$$x(t) = A\sin(\omega t + \phi) \tag{1}$$

where $A$ is the amplitude. The probability of finding the mass within an interval d$x$ at position $x$ is p($x$)d$x$, and this is equal to the fraction of time d$t$ that is spent in that interval over half a period. From Eq. (1) we readily determine that

$$dx = \omega A \cos(\omega t + \phi)\, dt = \omega\sqrt{A^2 - x^2}\, dt \tag{2}$$

resulting in the classical probability distribution

$$p_c(x) = \frac{1}{\pi\sqrt{A^2 - x^2}} = \frac{1}{\pi\sqrt{\frac{2E}{m\omega^2} - x^2}} \quad ; \quad x^2 \leq A^2$$
$$= 0 \quad ; \quad x^2 > A^2 \tag{3}$$

where we make use of the energy relation $E = \frac{1}{2}m\omega^2 A^2$.

The quantum situation is radically different. In place of Newton's equation of motion one uses the Schrödinger equation for the same potential. One now finds discrete energy levels

$$E_n = (n + \tfrac{1}{2})\hbar\omega . \tag{4}$$

The probability distribution for an oscillator in the $n$th excited state is

$$p_n(x) = \frac{\alpha}{\sqrt{\pi}2^n n!} H_n^2(\alpha x)\exp(-\alpha^2 x^2) \tag{5}$$

where $H_n(\alpha x)$ are the Hermite polynomials, and the scale factor $\alpha$ is

$$\alpha = \sqrt{\frac{m\omega}{\hbar}} . \tag{6}$$

For illustration we choose a quantum state excited to $n=12$ and compare $p_{12}(x)$ to the classical probability p$_c$(x). The amplitude of the latter has been set to unity by setting

$$E_n = \tfrac{1}{2}m\omega^2 = (n + \tfrac{1}{2})\hbar\omega \tag{7}$$

which results in $\alpha=5$ for $n=12$. These classical vs. quantum probability distributions are illustrated in Figure 1a. Classically, the mass spends more time near the extrema of the oscillation, where the velocity approaches zero, than at the center where the velocity is maximal.

For this degree of excitation, the quantum probability distribution oscillates around the classical distribution, and begins to approximate it in the mean.

The quantum zero-point state probability distribution is described by Eq. (5) with $n=0$, but in this case the expression may be simplified to

$$p_0(x) = \sqrt{\frac{m\omega}{\pi\hbar}} \exp(-m\omega x^2/\hbar) \ . \tag{8}$$

The classical distribution meanwhile remains the same on rescaling the axis. The two distributions are shown together in Figure 1b. They differ radically.

Any single mode of oscillation of the electromagnetic field will have analogous behavior to this simple harmonic oscillator. In section III we investigate the probability distribution for the zero-point field of quantum field theory and show that it has the form of Eq. (8). In section IV we verify that the Boyer classical representation of a single mode of the zero-point field has the form of Eq. (3), which is to be expected of a classical wave of definite amplitude (and therefore energy) but with an indefinite phase.

### III. QUANTUM FIELD THEORY FOR THE ZERO-POINT FIELD
#### A. Classical Probability Distributions for Quantum State Operators

We begin with an investigation of the precise correspondence between the statistics of the quantized radiation field, where the electric and magnetic fields are operators, and the statistics of a classical electromagnetic field consisting of plane waves forming the basis of the SED approach. To this end consider the relationship between a classical probability distribution and the expectation of an arbitrary function of a quantum operator (cf. Dirac [12]). Let us define $\hat{\mathbf{r}}$ to be some vector operator on a state $|\psi\rangle$, so that the expectation value of some function $f(\hat{\mathbf{r}})$ is $\langle\psi|f(\hat{\mathbf{r}})|\psi\rangle$. Provided that the components of the vector $\hat{\mathbf{r}}$ commute with one another, we may then associate with the particular state $|\psi\rangle$ a probability distribution (of a c-number $\mathbf{r}$) $p_\psi(\mathbf{r})$, say. The defining relation for this classical probability distribution is that the expectation value of an arbitrary function must be the same as that computed from quantum theory:

$$\langle\psi|f(\hat{\mathbf{r}})|\psi\rangle = \int d^M r \, p_\psi(\mathbf{r}) f(\mathbf{r}) \tag{9}$$

where $M$ is the dimensionality of $\mathbf{r}$. To determine $p_\psi(\mathbf{r})$ it is convenient to go via a generating function, $g(\mathbf{s})$ say, which is a transform of $p_\psi(\mathbf{r})$, and which can be computed directly from Eq. (9) if $f(\mathbf{r}) = f(\mathbf{r};\mathbf{s})$ is identified with the transform kernel. Choosing the Fourier transform,

$$f(\mathbf{r};\mathbf{s}) = \exp(-i\mathbf{s}.\mathbf{r}), \qquad (10)$$

one has:
$$g(\mathbf{s}) = \langle \psi | \exp(-i\mathbf{s}.\hat{\mathbf{r}}) | \psi \rangle. \qquad (11)$$

Once the generating function has been computed from Eq. (11), the inverse transform then gives the required distribution:

$$p_\psi(\mathbf{r}) = (2\pi)^{-M} \int d^M s\, g(\mathbf{s}) \exp(i\mathbf{s}.\mathbf{r}). \qquad (12)$$

In this paper we are interested exclusively in the ground state of the EM field $|\psi\rangle = |\mathbf{0}\rangle$, the distribution for which we denote simply by $p_\mathbf{0}(\mathbf{r}) = p(\mathbf{r})$.

### B. Probability Distribution for the Total Field

The electric field operator can be expanded in the photon number representation [14]

$$\hat{\mathbf{E}} = -i \sum_{\lambda=1}^{2} \sum_{\mathbf{k}} \boldsymbol{\varepsilon}_{\mathbf{k},\lambda} \sqrt{\frac{\hbar\omega}{2\varepsilon_0 V}} \left( \hat{a}^\dagger_{\mathbf{k},\lambda} \exp(-i\mathbf{k}.\mathbf{r} + i\omega t) - \hat{a}_{\mathbf{k},\lambda} \exp(i\mathbf{k}.\mathbf{r} - i\omega t) \right) \qquad (13)$$

where $\omega = ck$, $\boldsymbol{\varepsilon}_{\mathbf{k},\lambda}$ is a unit vector, and $\lambda$ labels the two directions mutually orthogonal and orthogonal to the momentum $\mathbf{k}$. Thus if each of the 3 Cartesian components of $\mathbf{k}$ can take N values, then the dimensionality of the representation space is 6N. In the following, unless $\lambda$ appears explicitly, sums and products involving $\mathbf{k}$ should be taken to include $\lambda$. In this paper we restrict our attention to the transverse components of the electric field, though the corresponding results for the magnetic field easily follow. Since all cavity modes are independent, the state of the total field can be written as the product of the individual modes (cf. [14], sect. 4.5). Specifically for the ground-state we may thus write

$$|\mathbf{0}\rangle \equiv \prod_{\mathbf{k}} |0_{\mathbf{k}}\rangle. \qquad (14)$$

We seek to relate a probability distribution for some function of the electric field regarded as a classical ensemble of waves to the measurement of that same function of the electric field regarded as a quantum operator; thus writing $\hat{\mathbf{E}} = (\hat{E}_x, \hat{E}_y, \hat{E}_z)$, and noting that the polarization components commute, a distribution p(**E**) is sought such that, with reference to Eq. (9), for an arbitrary function f,

$$\langle 0 | f(\hat{\mathbf{E}}) | 0 \rangle = \int d E_x \int d E_y \int d E_z \, p(\mathbf{E}) \, f(\mathbf{E}). \tag{15}$$

With reference to Eq. (11), the corresponding generating function is

$$g(\mathbf{s}) = \langle 0 | \exp(-i \mathbf{s} \cdot \hat{\mathbf{E}}) | 0 \rangle. \tag{16}$$

If g(**s**) can be calculated, then its Fourier inverse (Eq. (12)) gives p(**E**).

Substitution of Eq. (13) into the above permits the generating function to be written as a product:

$$g(\mathbf{s}) = \prod_{\mathbf{k}} \left\langle 0_{\mathbf{k}} \left| \exp\left[ -\mathbf{s} \cdot \boldsymbol{\varepsilon}_{\mathbf{k}} \sqrt{\frac{\hbar \omega}{2 \varepsilon_0 V}} \left( \hat{a}_{\mathbf{k}}^{\dagger} \exp(-i \mathbf{k} \cdot \mathbf{r} + i \omega t) - \hat{a}_{\mathbf{k}} \exp(i \mathbf{k} \cdot \mathbf{r} - i \omega t) \right) \right] \right| 0_{\mathbf{k}} \right\rangle. \tag{17}$$

Let us define a complex number $\alpha_{\mathbf{k}}$:

$$\alpha_{\mathbf{k}} \equiv -\mathbf{s} \cdot \boldsymbol{\varepsilon}_{\mathbf{k}} \sigma_{\mathbf{k}} \exp(-i \mathbf{k} \cdot \mathbf{r} + i \omega t) \tag{18}$$

where

$$\sigma_{\mathbf{k}}^2 \equiv \frac{\hbar \omega}{2 \varepsilon_0 V}. \tag{19}$$

Eq. (17) can now be written as

$$g(\mathbf{s}) = \prod_{\mathbf{k}} \langle 0_{\mathbf{k}} | \exp(\alpha_{\mathbf{k}} \hat{a}_{\mathbf{k}}^{\dagger} - \alpha_{\mathbf{k}}^{*} \hat{a}_{\mathbf{k}}) | 0_{\mathbf{k}} \rangle. \tag{20}$$

We now recognize a connection between Eq. (20) and the coherent photon state, $|\alpha_{\mathbf{k}}\rangle$ say. This can be written in terms of an operation on the ground state (cf. [14] sect. 4.11)

$$|\alpha_{\mathbf{k}}\rangle = \exp(\alpha_{\mathbf{k}} \hat{a}_{\mathbf{k}}^{\dagger} - \alpha_{\mathbf{k}}^{*} \hat{a}_{\mathbf{k}}) | 0_{\mathbf{k}} \rangle \tag{21}$$

which permits Eq. (17) to be written

$$g(\mathbf{s}) = \prod_{\mathbf{k}} \langle 0_{\mathbf{k}} | \alpha_{\mathbf{k}} \rangle. \tag{22}$$

However, the basis states for the coherent state are well known:

$$|\alpha\rangle = \exp(-|\alpha|^2 / 2) \sum_{n} \frac{\alpha^n}{\sqrt{n!}} |n\rangle. \tag{23}$$

This allows us to arrive immediately at

$$g(\mathbf{s}) = \exp\left( -\sum_{\mathbf{k}} |\alpha_{\mathbf{k}}|^2 / 2 \right). \tag{24}$$

If the space is unbounded, then the sum over **k** in Eq. (24) can be replaced:

$$\sum_{\mathbf{k}} \rightarrow \frac{V}{(2\pi c)^3} \int_0^\infty d\omega\, \omega^2 \sum_{\lambda=1}^{2} \int d\Omega_{\mathbf{k}} \tag{25}$$

(cf. [14], sect. 1.1, noting the integration over $4\pi$ solid angle and summation over both polarization states) and the integral over orientations can be performed:

$$\sum_{\lambda=1}^{2} \int d\Omega_{\mathbf{k}} (\mathbf{s}.\boldsymbol{\varepsilon}_{\mathbf{k},\lambda})^2 = \frac{8\pi s^2}{3}. \tag{26}$$

Substitution of Eqs. (25) and (26) into Eq. (24) gives

$$g(\mathbf{s}) = \exp(-s^2 \sigma_E^2 / 2) \tag{27}$$

where

$$\sigma_E^2 \equiv \frac{\hbar \int d\omega\, \omega^3}{6\pi^2 \varepsilon_0 c^3}. \tag{28}$$

Finally, performing the Fourier inverse prescribed by Eq. (12) gives

$$p(\mathbf{E}(\mathbf{r},t)) = (2\pi\sigma_E^2)^{-3/2} \exp[-E^2(\mathbf{r},t)/2\sigma_E^2]. \tag{29}$$

where it has been made explicit that this is the distribution for the total field at any point in (*r*,*t*). Thus the probability distribution for an unbounded zero-point field is isotropic and normal with zero mean. In the absence of a cutoff frequency the variance is infinite, and therefore the distribution is effectively flat, each component of the polarization vector having an amplitude that is equally likely to take any value.

### C. Zero-point Energy of the Quantum Field

The following is a brief example of how a quantum mechanical expectation value can be replaced by a classical average. The expected value for the zero-point energy, assuming equal contributions from the electric and magnetic field components is

$$\mathcal{E}_{zp} = \varepsilon_0 V \langle 0|\hat{\mathbf{E}}.\hat{\mathbf{E}}|0\rangle. \tag{30}$$

According to Eq. (15), this may now be replaced by

$$\mathcal{E}_{zp} = \varepsilon_0 V \int dE_x \int dE_y \int dE_z\, p(\mathbf{E}) E^2. \tag{31}$$

Using the result Eq. (29), and performing the integrals, one obtains

$$\mathcal{E}_{zp} = 3\varepsilon_0 V \sigma_E^2 = \frac{V\hbar \int d\omega\, \omega^3}{2\pi^2 c^3} \tag{32}$$

which gives an energy density (per radian frequency, per unit volume)

$$\rho(\omega) = \frac{\hbar \omega^3}{2\pi^2 c^3} \qquad (33)$$

which is the required (QFT) density corresponding to average energy $\frac{1}{2}\hbar\omega$ per normal mode.

### D. Probability distribution of a single mode

A distribution for the amplitude of each normal mode can be obtained by decomposing the electric field operator into a sum of commuting normal mode operators

$$\hat{\mathbf{E}} = \sum_{\mathbf{k}} \boldsymbol{\varepsilon}_{\mathbf{k}} \hat{E}_{\mathbf{k}} . \qquad (34)$$

Following the same procedure as for the full field, it is straightforward to show that each normal mode amplitude is itself normally distributed:

$$p(E_{\mathbf{k}}) = \frac{1}{\sqrt{2\pi\sigma_{\mathbf{k}}^2}} \exp(-E_{\mathbf{k}}^2 / 2\sigma_{\mathbf{k}}^2) . \qquad (35)$$

where $\sigma_{\mathbf{k}}$ is as defined in Eq. (19). This distribution for the amplitude of the electromagnetic field modes is identical to that of the quantum mechanical simple harmonic oscillator of Eq. (8) as illustrated in Figure 1b. (for $\frac{1}{2}m\omega^2 = \varepsilon_0 V$). In this case, as the space becomes unbounded, the variance becomes vanishingly small, and the distribution approaches a delta function centered on an amplitude value of zero. However, the zpf energy in each mode, $\mathcal{E}_{\mathbf{k}}$ say, remains finite and independent of the geometry:

$$\mathcal{E}_{\mathbf{k}} = \varepsilon_0 V \int dE_{\mathbf{k}} \, p(E_{\mathbf{k}}) E_{\mathbf{k}}^2 = \tfrac{1}{2}\hbar\omega \qquad (36)$$

as required.

### IV. BOYER'S CLASSICAL STOCHASTIC ZERO-POINT FIELD
### A. Introduction

The question we seek to address is whether it is possible to substitute a classical electromagnetic field with some stochastic behavior for the zero-point field of QFT. Such a field would be expected to correctly predict effects attributed to the zero point field, including spontaneous emission, Van-der-Waals forces, the Lamb shift, and Casimir forces. Boyer's suggestion [1] is a classical field which is a sum of Fourier **k** components of definite amplitude, but random phase. Averaging over the phases then replaces the procedure of taking the ground-state expectation value of an operator involving the electromagnetic field.

To be consistent with the notation of this paper, we write the Boyer field as:

$$\mathbf{E}(\mathbf{r},t) = \sqrt{2}\sum_{\mathbf{k}} \boldsymbol{\varepsilon}_{\mathbf{k}} \sigma_{\mathbf{k}} \cos(\mathbf{k}\cdot\mathbf{r} - \omega t + \theta_{\mathbf{k}}) \tag{37}$$

with the $\theta_{\mathbf{k}}$ independent and subject to a constant distribution in $[0, 2\pi]$:

$$\langle \exp(i\theta_{\mathbf{k},\lambda} - i\theta_{\mathbf{k}',\lambda}) \rangle = \delta_{\mathbf{k},\mathbf{k}'} \tag{38}$$

That this is equivalent to the Boyer field of earlier publications can be shown by converting the sum to an integral (valid when the space is unbounded), which from Eq. (25) is

$$\sum_{\mathbf{k}} \rightarrow \frac{V}{(2\pi)^3} \int_0^\infty d^3k \sum_{\lambda=1}^{2}, \tag{39}$$

and at the same time redefining the random variables so that

$$\langle \exp(i\theta_{\mathbf{k},\lambda} - i\theta_{\mathbf{k}',\lambda}) \rangle = \delta(\mathbf{k}-\mathbf{k}')\delta_{\lambda,\lambda'}. \tag{40}$$

Applying Eq. (39) to (37) gives the expression used in Boyer [1] (but in MKS units):

$$\mathbf{E}(\mathbf{r},t) = \text{Re}\sum_{\lambda=1}^{2}\int d^3k\, \boldsymbol{\varepsilon}_{\mathbf{k}} \sqrt{\frac{\hbar\omega}{8\pi^3\varepsilon_0}} \exp(i\mathbf{k}\cdot\mathbf{r} - i\omega t + i\theta_{\mathbf{k}}). \tag{41}$$

**B. Distribution for the total field**

A requirement of the classical stochastic field is that it have the same distribution as the QFT zero-point field. The generating function for the distribution is defined in the normal way, but now the expectation value implies an averaging over the phases:

$$\begin{aligned}
g(\mathbf{s}) &\equiv \langle \exp(-i\mathbf{s}\cdot\mathbf{E}) \rangle \\
&= \prod_{\mathbf{k}} \left\{ \frac{1}{2\pi} \int_0^{2\pi} d\theta_{\mathbf{k}} \exp\left(-i\sqrt{2}\sigma_{\mathbf{k}} \mathbf{s}\cdot\boldsymbol{\varepsilon}_{\mathbf{k}} \cos(\mathbf{k}\cdot\mathbf{r}-\omega t + \theta_{\mathbf{k}})\right) \right\} \\
&= \prod_{\mathbf{k}} \left\{ J_0(\sqrt{2}\sigma_{\mathbf{k}} \mathbf{s}\cdot\boldsymbol{\varepsilon}_{\mathbf{k}}) \right\}
\end{aligned} \tag{42}$$

which is just the generating function for the sum of a set of independent classical harmonic oscillators. In the limit that $\sigma_{\mathbf{k}}$ is small (the volume is large), the Bessel functions are always positive, and it is safe to write:

$$g(\mathbf{s}) = \exp\left(\sum_{\mathbf{k}} \log\left(J_0\left(\sqrt{2}\sigma_{\mathbf{k}} \mathbf{s}.\boldsymbol{\varepsilon}_{\mathbf{k}}\right)\right)\right) \qquad (43)$$

The generating function for the QFT distribution, Eq. (24), can be reproduced if only the first two terms are retained in the power series expansion of the Bessel function, and provided one can then set $\log(1+x) \approx x$ for each term in the sum, whereupon

$$g(\mathbf{s}) \approx \exp\left(-\sum_{\mathbf{k}} |\alpha_{\mathbf{k}}|^2/2\right) \qquad (44)$$

where $\alpha_{\mathbf{k}}$ is defined in Eq. (18). The above result is valid in the limit that the volume is infinite, because the sum over $\mathbf{k}$ supplies a volume which cancels with that in $\alpha_{\mathbf{k}}$. It follows that higher order terms ($n>1$) in the expansion of the Bessel and log functions must have dependencies on the volume which go as $V^{1-n}$. Therefore, when the volume is bounded (in any dimension, such as for example by a Casimir cavity) then this classical distribution will differ from the QFT distribution. (Note that within the QFT framework the generating function Eq. (24) is exact, regardless of the volume of the space supporting the field modes.)

### C. Distribution for a single mode

The disagreement between QFT and the Boyer field is more acute for a single mode. The QFT result is that *each mode* is independently normally distributed. The Boyer field components are also independent, but instead have the distribution of a classical harmonic oscillator, i.e. each component of the expansion

$$\mathbf{E} = \sum_{\mathbf{k}} \boldsymbol{\varepsilon}_{\mathbf{k}} E_{\mathbf{k}} \qquad (45)$$

has the distribution given in Eq. (3), which in the notation of this section is

$$\begin{aligned} p(E_{\mathbf{k}}) &= \frac{1}{\pi\sqrt{\sigma_{\mathbf{k}}^2 - E_{\mathbf{k}}^2}} & E_{\mathbf{k}}^2 &\leq \sigma_{\mathbf{k}}^2 \\ &= 0 & E_{\mathbf{k}}^2 &> \sigma_{\mathbf{k}}^2 \end{aligned} \qquad (46)$$

Comparison with the QFT result Eq. (35) will show that the first and second moments are the same, but beyond that the distributions diverge widely. The conclusion is that a classical field with a random phase does not accurately reproduce the statistics of a QFT field operator on the ground-state.

## V. A MODIFIED CLASSICAL ZERO-POINT FIELD

### A. Introduction

An alternative to the Boyer field is the following:

$$\mathbf{E}(\mathbf{r},t) = \text{Re} \sum_{\mathbf{k}} \boldsymbol{\varepsilon}_{\mathbf{k}} \sigma_{\mathbf{k}} w_{\mathbf{k}} \exp(i\mathbf{k}\cdot\mathbf{r} - i\omega t) \qquad (47)$$

where $w_{\mathbf{k}} = u_{\mathbf{k}} + iv_{\mathbf{k}}$, and $u_{\mathbf{k}}$ and $v_{\mathbf{k}}$ are real, independent, normally distributed random variables, having zero mean and unit variance. This field can alternatively be written:

$$\mathbf{E}(\mathbf{r},t) = \sqrt{2} \sum_{\mathbf{k}} \boldsymbol{\varepsilon}_{\mathbf{k}} \sigma_{\mathbf{k}} \sqrt{I_{\mathbf{k}}} \cos(\mathbf{k}\cdot\mathbf{r} - \omega t + \theta_{\mathbf{k}}) \qquad (48)$$

where $\theta_{\mathbf{k}}$ has a constant distribution in $[0, 2\pi]$, and $I_{\mathbf{k}}$ has distribution:

$$p(I_{\mathbf{k}}) = \exp(-I_{\mathbf{k}}); \qquad I_{\mathbf{k}} \in [0, \infty]. \qquad (49)$$

i.e. the intensity of each mode is exponentially distributed. Comparison of Eq. (48) with Eq. (37) indicates that one may regard this field as differing from the Boyer field by the introduction of a stochastic element for the amplitude of each mode. (The stochastic nature of this field can be viewed as an implementation of the Box-Muller method for generating normally distributed random deviates from a pair of uniformly distributed random deviates, c.f. [15].)

### B. Distribution of a single mode

Expanding the field as in Eq. (37) it is seen that the amplitude $E_{\mathbf{k}}$ of each mode is a linear combination of two independent zero mean unit normal deviates:

$$E_{\mathbf{k}} = \sigma_{\mathbf{k}} \left( u_{\mathbf{k}} \cos(\mathbf{k}\cdot\mathbf{r} - \omega t) - v_{\mathbf{k}} \sin(\mathbf{k}\cdot\mathbf{r} - \omega t) \right). \qquad (50)$$

It follows that $E_{\mathbf{k}}$ is itself normally distributed with variance

$$<E_{\mathbf{k}}^2> = \sigma_{\mathbf{k}}^2 \left[ <u_{\mathbf{k}}^2> \cos^2(\mathbf{k}\cdot\mathbf{r} - \omega t) + <v_{\mathbf{k}}^2> \sin^2(\mathbf{k}\cdot\mathbf{r} - \omega t) \right] = \sigma_{\mathbf{k}}^2 \qquad (51)$$

from which follows the required result that each component is distributed in accord with Eq. (35). Since the stochastic variables are defined to be independent, and since each mode of this field has the same distribution as the modes of the QFT zero point field, it follows that the distributions for the full fields also agree. The conclusion is that the field defined in Eq. (47) is the correct classical analogue of the QFT zero point field.

### C. Harmonic oscillator driven by a stochastic field

Stochastic electrodynamics attempts to use a stochastic but otherwise classical field to predict the motion of charged particles as given by quantum theory. A recent success has been the prediction of the correct (RMS) radius for the ground-state of the Bohr atom by interpreting the Boyer field as a forcing term on a classical harmonic oscillator [1]. That analysis is extended here using the alternative classical zero-point field presented above to determine the full probability distribution for the oscillator coordinate.

Using the notation of Puthoff [7], the equation of motion for a classical electron oscillator driven by the zero-point field, including radiation damping, is:

$$\ddot{\mathbf{q}} + v_0^2 \mathbf{q} - \Gamma \dddot{\mathbf{q}} = \Gamma' \mathbf{E} \tag{52}$$

where $\mathbf{q} \equiv \mathbf{q}(t)$ is the oscillator co-ordinate, $v_0$ is the natural frequency of the oscillator, and the damping and driving coefficients are:

$$\Gamma = e^2 / 6\pi\varepsilon_0 m_e c^3, \quad \Gamma' = e / m_e. \tag{53}$$

Implicit in Eq. (52) is that the velocities are non-relativistic, whence the field can be treated as if acting at a constant point, which for convenience is taken to be at $\mathbf{r}=0$. Using the Fourier transforms of the coordinate and the field, Eq. (52) can be written

$$\mathbf{q}(v) = \mathrm{h}(v)\mathbf{E}(v)$$
$$\mathrm{h}(v) \equiv \frac{\Gamma'}{v_0^2 - v^2 + i\Gamma v^3}. \tag{54}$$

The Fourier transform of the field given by Eq. (47) (suppressing the argument $\mathbf{r}=0$) is

$$\mathbf{E}(v) = \pi \sum_\mathbf{k} \sigma_\mathbf{k} \boldsymbol{\varepsilon}_\mathbf{k} \left[ \delta(\omega + v)(u_\mathbf{k} + iv_\mathbf{k}) + \delta(\omega - v)(u_\mathbf{k} - iv_\mathbf{k}) \right]. \tag{55}$$

Substitution of Eq. (55) into the Fourier inverse of $\mathbf{q}(v)$ (as given in Eq. (54)) gives

$$\mathbf{q}(t) = \int dv \exp(ivt)\, \mathrm{h}(v)\mathbf{E}(v) / 2\pi$$
$$= \sum_\mathbf{k} \boldsymbol{\varepsilon}_\mathbf{k} \sigma_\mathbf{k} \operatorname{Re}\{ w_\mathbf{k} \widetilde{\mathrm{h}}^*(\omega) \}, \tag{56}$$
$$\widetilde{\mathrm{h}}(\omega) \equiv \exp(i\omega t)\, \mathrm{h}(\omega).$$

The generating function for the distribution is

$$g(\mathbf{s}) \equiv \langle \exp(-i\mathbf{s}.\mathbf{q})\rangle$$
$$= \prod_k \left\{\frac{1}{2\pi}\int du_k \int dv_k \exp\left[-u_k^2/2 - v_k^2/2 - i\mathbf{s}.\boldsymbol{\varepsilon}_k \sigma_k \left(u_k \operatorname{Re}\{\tilde{h}(\omega)\} + v_k \operatorname{Im}\{\tilde{h}(\omega)\}\right)\right]\right\} \quad (57)$$
$$= \prod_k \exp\left(-|\mathbf{s}.\boldsymbol{\varepsilon}_k \sigma_k h(\omega)|^2/2\right).$$

This result is exact: no approximation based upon the geometry of the bounding space, and therefore the mode density, has been made thus far. The generating function, and therefore the distribution of the oscillator coordinate, can in principle be computed from the above for any set of modes associated with a given (finite) geometry. Stochastic electrodynamics therefore predicts a dependency of the oscillator amplitude on the geometry of the space supporting the zero point field. (It has not yet been shown if this dependency agrees with that of a quantum-theoretical model which takes into account the reflective properties of the confining walls on the harmonic potential.)

In an unbounded space the product may be converted into an integral of the exponent, as in the earlier analysis. Using Eqs. (25) and (26), one thereby obtains

$$g(\mathbf{s}) = \exp\left(-\frac{s^2 \hbar}{12\pi^2 c^3 \varepsilon_0}\int_0^\infty d\omega\, \omega^3 |h(\omega)|^2\right). \quad (58)$$

Using the resonance approximation [7], the integral is found to be

$$\int_0^\infty d\omega\, \omega^3 |h(\omega)|^2 \approx \frac{\pi(\Gamma')^2}{2\Gamma\nu_0} = \frac{3\pi^2 c^3 \varepsilon_0}{m_e \nu_0}. \quad (59)$$

Inserting this result into Eq. (58) gives

$$g(\mathbf{s}) = \exp(-s^2 \sigma_q^2/2) \quad (60)$$

where
$$\sigma_q^2 = \frac{\hbar}{2m_e \nu_0} \quad (61)$$

whence the distribution for the oscillator coordinate is

$$p(\mathbf{q}) = (2\pi\sigma_q^2)^{-3/2} \exp(-q^2/2\sigma_q^2). \quad (62)$$

This distribution agrees with that predicted by quantum mechanics for the non-relativistic harmonic oscillator in the ground-state.

### D. Correspondence with the Bohr atom

The Bohr atom (ground-state) can be considered as a pair of 1D harmonic oscillators [1,7], oscillating in quadrature in a plane. Confining the distribution Eq. (62) to 2 dimensions gives

$$p(q_x, q_y) = \left(2\pi\sigma_\mathbf{q}^2\right)^{-1} \exp\left(-\frac{q_x^2 + q_y^2}{2\sigma_\mathbf{q}^2}\right). \qquad (63)$$

from which the Bohr radius is predicted to be

$$<r_0^2> = <q_x^2 + q_y^2> = 2\sigma_\mathbf{q}^2 = \frac{\hbar}{m_e v_0} \qquad (64)$$

which is the required result according to quantum theory. No attempt is made here to replicate the quantum-mechanical *distribution* of the radius for an atom in the ground-state because the correspondence between the classical and quantum models is incomplete. A fuller treatment requires first a solution for the equation of motion for the position of a classical electron in a $1/r^2$ force-field including radiation damping and the force from a stochastic zero-point field.

## VI. DISCUSSION

Second quantization of the electromagnetic field leads to a system of Schrödinger type equations for the QFT 'wave-function' in a space whose coordinates are the amplitudes of the Fourier expansion of the vector potential. The potential experienced by this wavefunction is the energy associated with the amplitudes of the Fourier modes, i.e. is proportional to the square of the amplitude. Thus each mode contributes a one-degree-of-freedom harmonic oscillator to the dynamics. It follows that the QFT wavefunction for a single mode is in general a super-position of the Hermite-Gauss functions of the Fourier amplitude, the ground-state of which is just the Gaussian. Therefore the probability density for the amplitude of a mode when the system is in the ground-state is also a Gaussian. It is not surprising therefore that the ground-state *electric field* amplitude is also normally distributed, as shown in section III.

In section IV the Boyer field has been shown to replicate the distribution of the total field only by virtue of the mean value theorem. The individual modes are not normally distributed, but conspire to give a normal distribution to the amplitude of the full field in the limit of an infinite density of **k** states. The deployment of the Boyer field as the SED driving force in a classical harmonic oscillator succeeds in reproducing the correct RMS amplitude provided the linewidth (Γ in Eqs. (52) and (53)) is sufficiently broad - i.e. provided there are a sufficient number of field modes within the bandwidth centered on the resonant frequency of the oscillator.

The modified stochastic character of the field introduced in section V does not suffer from this limitation. Rather, it has been shown that a classical EM field with a suitable stochastic behavior not only reproduces the statistics of the zero-point field of Quantum Field Theory, but also imposes the correct - quantum mechanical - distribution upon a *classical* harmonic oscillator embedded in such a field. However, the aim of this paper is not to simply provide a more accurate classical analogue for the zero-point field. Rather, we wish to draw attention to the connection between the probability distributions of the stochastic variables in the classical field, and those of the Fourier amplitudes of the second quantized field theory. It is evident that in order to accurately reproduce QFT statistics, the classical field must borrow the appropriate distributions from QFT. In future work we hope to show how this procedure can be carried further, and derive a classical field with a stochastic character that successfully imitates elevated and mixed states of the quantized field.

## ACKNOWLEDGMENTS

Michael Ibison gratefully acknowledges support from the McDonnell Foundation Inc. and the Human Information Processing Group at Princeton University.

## REFERENCES

[1]  T.H. Boyer, Phys. Rev. D, **11**, 790 (1975).

[2]  P.W. Milonni, *The Quantum Vacuum: An Introduction to Quantum Electrodynamics*, (Academic Press) (1994).

[3]  M. Planck, Verh. Dtsch. Phys. Ges. **13**, 138 (1911); Ann. Phys. (Leipzig) **37**, 642 (1912).

[4]  W. Nernst, Verh. Dtsch. Phys. Ges. **18**, 83 (1916).

[5]  A. Einstein and O. Stern, Ann. Phys. (Leipzig) **40**, 551 (1913).

[6]  T.H. Boyer, Phys. Rev. D, **29**, 1096 (1984).

[7]  H.E. Puthoff, Phys. Rev. D **35**, 3266 (1987).

[8]  A.D. Sakharov, Dokl. Akad. Nauk SSSR [Sov. Phys. Dokl. **12**, 1040 (1968)].

[9]  H.E. Puthoff, Phys. Rev. A, **39**, 2333 (1989).

[10]  B. Haisch, A. Rueda and H.E. Puthoff, Phys. Rev. A, **49**, 678 (1994).

[11]  D.P. Dutta, Class. Quantum Grav., **12**, 2499 (1995).

[12]  P.C.W. Davies and D.S. Betts, *Quantum Mechanics* (2nd edition), (Chapman and Hall), pp. 37-42 (1994).

[13]  P.A.M. Dirac, *The Principles of Quantum Mechanics*, (Oxford), pp. 45-48 (1958).

[14]  R. Loudon, *The Quantum Theory of Light,* (2nd edition) (Oxford University Press), Oxford, UK, (1983).

[15]  L. Devroye, *Non-Uniform Random Variate Generation*, (Springer-Verlag), New York, §9.1 (1986).